\documentclass[conference]{sig-alternate-10pt}
\usepackage{amsmath, amssymb}
\usepackage{graphicx}
\usepackage{color}
\usepackage{multirow}
\usepackage{abbrevs}
\usepackage{ifpdf}
\usepackage{graphicx}
\usepackage{epstopdf}
\usepackage{cite}
\usepackage{amsfonts}
\usepackage[tight,footnotesize]{subfigure}

\usepackage{marginnote}

\begin{document}

\title{RASID: A Robust WLAN Device-free Passive Motion Detection System}

\numberofauthors{3}
\author{
\alignauthor
Ahmed E. Kosba\\
       \affaddr{Dept. of Comp. and Sys. Eng.}\\
       \affaddr{Faculty of Engineering}\\
       \affaddr{Alexandria University, Egypt}\\
       \email{ahmed.kosba@alexu.edu.eg}
% 2nd. author
\alignauthor
Ahmed Saeed\\
      \affaddr{Dept. of Comp. Sc. and Eng.}\\
       \affaddr{Egypt-Japan Univ. of Sc. and Tech. (E-JUST), Egypt}\\
%       \affaddr{Egypt}\\
       \email{ahmed.saeed@ejust.edu.eg}
\alignauthor
Moustafa Youssef\\
       \affaddr{Dept. of Comp. Sc. and Eng.}\\
       \affaddr{Egypt-Japan Univ. of Sc. and Tech. (E-JUST), Egypt}\\
 %      \affaddr{Egypt}\\
       \email{moustafa.youssef@ejust.edu.eg}
}

\maketitle

\begin{abstract}

WLAN Device-free passive (\textit{DfP}) indoor localization is an
emerging technology enabling the localization of entities that do
not carry any devices nor participate actively in the localization
process using the already installed wireless infrastructure. This
technology is useful for a variety of applications such as intrusion
detection, smart homes and border protection.

We present the design, implementation and evaluation of \textit{RASID},
a \textit{DfP} system for human motion detection.
\textit{RASID}
combines different modules for statistical anomaly detection while
adapting to changes in the environment to provide accurate, robust,
and low-overhead detection of human activities using standard WiFi hardware. Evaluation of the
system in two different testbeds shows that it can
achieve an accurate detection capability in both environments with an F-measure of at least 0.93.
In addition, the high accuracy and low overhead performance are robust to changes in the
environment as compared to the current state of the art \emph{DfP}
detection systems.
We also relay the lessons learned during building
our system and discuss future research directions.

\end{abstract}

\keywords{Anomaly detection, device-free passive localization,
motion detection systems, robust device-free localization.}

\section{Introduction}

The increasing need for context-aware information and the rapid advancements in communication networks have motivated significant research effort in the area of location-based services. This effort resulted in the development of many location determination systems, including the GPS system \cite{GPS}, ultrasonic-based systems \cite{ultrasonic}, infrared-based (IR) systems \cite{infrared}, and radio frequency-based (RF) systems \cite{Horus}. Moreover, motion detection systems, that aim at detecting the motion of an entity carrying a device, were also developed \cite{SmartMoveX, ContextAwareness, ActivityRecognition, Inferring, WStationMotion, Locadio, Intraroom, GSM1, GSM2}. These systems require the tracked entity to carry a device that participates in the localization process. Thus, we refer to them as device-based systems.

\begin{figure}[!t]
    \centering
        \includegraphics[width=0.5\textwidth]{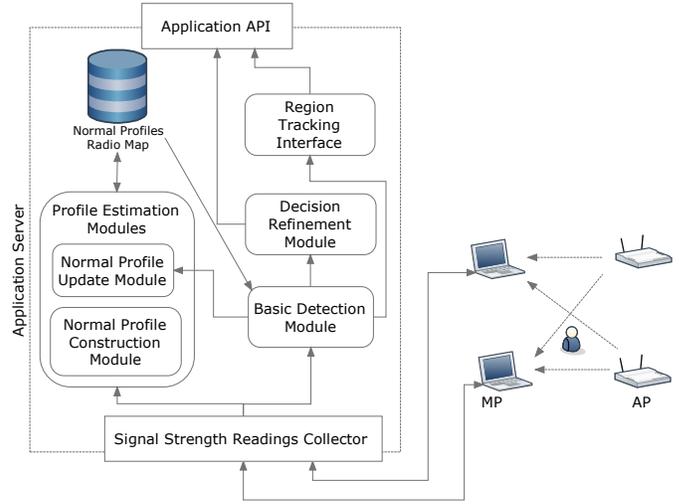}
    \caption{\textit{RASID} system architecture.}
    \label{fig:arch}
\end{figure}

Motivated by the wide use of wireless LANs for indoor communication,
we recently introduced the concept of device-free passive
\textit{DfP} localization \cite{Challenges} which enables the detection and tracking of entities that do not carry any devices nor
participate in the localization process. This concept depends on the
fact that the presence and motion of entities in an RF environment
affects the RF signal strength, especially when dealing with the 2.4
GHz band which is used in different IEEE standards such as 802.11b
and 802.11g (WiFi).
Different \textit{DfP} algorithms were proposed for detection
\cite{Challenges, Smart} and tracking \cite{Challenges, Analysis,
Nuzzer, Kalman} of entities in indoor environments. Our focus in this paper is on the detection problem.

In particular, we address the problem of designing a low-overhead,
accurate, and robust \textit{DfP} motion detection system. We
introduce the \textit{RASID} system that provides a software only solution on top of the already installed wireless networks
enabling a wide set of applications including intrusion detection, border protection, and smart homes.
As a typical \emph{DfP} system, \textit{RASID} consists of
signal transmitters, such as access points (APs), signal receivers
or monitoring points (MPs), such as standard laptops\footnote{Note that it is also possible to use the access points themselves as monitoring points.}, and
an application server which collects and processes information about
the received signals from each MP. The application server contains the main system modules
responsible for performing the detection function (Figure \ref{fig:arch}).

Our research on \textit{RASID} is motivated by several factors: First, the technologies that can be used to provide the desired detection capability (e.g. cameras \cite{compVision}, IR sensors, radio tomographic imaging \cite{RTI}, pressure sensors \cite{phyContact}, etc) share the requirement of installing special hardware. In addition, cameras and IR sensors are limited
to line-of-sight vision and thus the cost of covering an area might be prohibitive.
Moreover, regular cameras fail to work in the dark or in the presence
of smoke. %, and they can cause privacy concerns.
\textit{RASID} avoids these drawbacks by using the already installed wireless infrastructure without installing any special hardware. It also makes use of the fact that RF waves do not require LOS for propagation.

From another perspective, the previously proposed WLAN \emph{DfP} detection techniques \cite{Smart,Challenges} provide good performance under strong assumptions, which limit their application domain. For
example, they are not robust to changes in the
environment. That is they do not adapt to changes in
the environment, e.g. humidity and temperature changes. Moreover, their parameters need to be changed as the deployment area changes.
In addition, the technique proposed in \cite{Smart} requires the construction of a human motion profile which leads to high overhead inside large-scale environments. The cost of this technique may be prohibitive, as it requires access to all areas of a
building which might include restricted or private areas and
requires several hours of calibration. Finally, all techniques were either
evaluated in controlled environments, e.g.
\cite{Challenges}, or in small-scale real
environments, e.g. \cite{Smart}.

In order to achieve its objectives, \textit{RASID} uses a statistical anomaly
detection technique to detect motion inside indoor environments.% \cite{RASID_WCNC}.
\textit{RASID} only constructs a non-parametric profile for the signal strength readings received
at the MPs when there is no human activity during a short
training phase of only two minutes, leading to minimal deployment overhead.
\textit{RASID} also employs techniques for continuously updating its silence profile to adapt to the environment changes. The system also applies a decision refinement procedure in order to reduce the false alarms due to the signal noise.
Furthermore, \textit{RASID} also provides an interface by which the regions of activity can be identified.
We evaluate the system in two different large-scale environments
rich in multi-path and compare \textit{RASID} to the state-of-the-art
\textit{DfP} detection techniques \cite{Challenges, Smart}. Our results show that
\textit{RASID} achieves its goals of high accuracy in both environments with
minimal deployment overhead. In addition, it is
robust to changes in the environment.

In summary, the contributions of this paper are four-fold:
\begin{itemize}
\item We present the architecture and implementation of \textit{RASID}: a system that provides robust device-free motion detection along with techniques for adapting to environment changes and handling the wireless signal noise. 
	\item We analyze different signal strength features that can be used for detection and identify the most promising one.
	\item We evaluate the system in two different large-scale real testbeds and compare it to the state-of-the-art \emph{DfP} detection techniques.
	\item We present a comparison of parametric and non-parametric approaches for system operation.
\end{itemize}

The rest of this paper is organized as follows: Section \ref{sec:RelatedWork} reviews related work. Section \ref{sec:RASID} presents the \textit{RASID} system architecture and operation. Section \ref{sec:Evaluation} presents the experimental evaluation of \textit{RASID} and a comparison with other
techniques. Section \ref{sec:analyticalmodel} compares the non-parametric approach used in the system to a parametric analytical model for the system operation. In Section \ref{sec:Discussion}, we discuss our experience with \textit{RASID} and
present some open research issues for future work. Finally, Section \ref{sec:Conclusion} concludes the paper and discusses future work.

\section{Related Work}
\label{sec:RelatedWork}

Motion Detection in \textbf{device-based} systems has been an active
field of research. Several works have been proposed to detect the
motion of an entity carrying a device either with the use of special
hardware like accelerometers or motion sensors \cite{SmartMoveX,
ContextAwareness, ActivityRecognition, Inferring}, or by using the
existing network infrastructures like wireless networks
\cite{WStationMotion, Locadio, Intraroom} and GSM \cite{GSM1, GSM2}.

From the \textbf{device-free} perspective, multiple technologies can
be used to provide the desired capabilities including: ultra-wide
band radar \cite{radar2}, computer vision \cite{compVision},
physical contact based systems \cite{phyContact} and radio
tomographic imaging \cite{RTI}. Other technologies include the usage
of wireless sensors for tracking transceiver-free objects
\cite{Transceiver-free} as well as the usage of RFID tags
\cite{Geometric}. Those technologies share the requirement of
installing special hardware to handle the device-free different
functionalities. In addition, cameras and IR sensors are limited to
line-of-sight vision and thus they require a high cost deployment to
cover all site regions. Moreover, regular cameras can fail to work
in the dark or in the presence of smoke, and they can cause privacy
concerns. Ultra-wide band radar based techniques also suffer from
high complexity. Moreover, some techniques can require high density
to provide full coverage like radio tomographic imaging and physical
contact based systems using pressure sensors.

WLAN device-free passive systems try to avoid the above drawbacks by
using the already available wireless infrastructure. The concept of
device-free passive detection and tracking using WLANs was first
proposed in \cite{Challenges} with a large number of applications including intrusion detection, border protection \cite{CyPhyCARD}, smart homes, and traffic estimation\cite{ReVISE_VTC}. Techniques for \textit{DfP} detection
\cite{Challenges, Smart} and tracking \cite{Challenges, Analysis,
Nuzzer} were introduced. The proposed techniques for the detection
capability are either based on time-series analysis like the moving
average and moving variance techniques proposed in \cite{Challenges}
or based on classification using the maximum likelihood estimation
\cite{Smart}.

In comparison, \textit{RASID} uses anomaly detection techniques to
identify the deviations from the normal (silence) state.
\textit{RASID} system uses a semi-supervised statistical technique
that models the learned normal behavior using a kernel-function
based non-parametric estimation. The kernel-function based anomaly
detection has been used in several applications where the
distribution of the normal behavior is not known. For example,
non-parametric estimation using Gaussian kernels was used in network
intrusion detection \cite{ParzenDetectors} and novelty detection
applied to oil flow data \cite{Novelty}. Also, density estimation
using Epanechnikov kernels was used in online outlier detection in
sensor data \cite{OnlineSensor} and to achieve continuous adaptive
outlier detection on distributed data streams
\cite{AdaptiveStreams}.

Compared to the previously proposed WLAN \textit{DfP} detection
techniques, the usage of the statistical anomaly detection
technique, along with the other techniques devised for adapting to
environment changes and refining the decision, enable \textit{RASID}
to achieve low deployment overhead, high accuracy and high
robustness.

\section{The RASID System}
\label{sec:RASID} In this section, we give the details of the
\textit{RASID} system. We start by an overview of the system
architecture followed by the details of the system modules.

\subsection{System Overview}
Figure~\ref{fig:arch} gives an overview of the system architecture.
The modules of the proposed system are implemented in the
application server that collects samples from the monitoring points
and processes them.  The system works in two phases: 1) A short
\emph{offline} phase, during which the system studies the signal
strength values when no human is present inside the area of interest
to construct what we call a ``normal or silence
profile'' for each stream. The profiles of all streams are constructed concurrently in that short phase.
2) A \emph{monitoring} phase, in which the system
collects readings from the monitoring points and decides whether
there is human activity (anomalous behavior) or not based on the
information gathered in the offline phase. It also updates the
stored normal profile so that it can adapt to environment changes.
Finally, a decision refinement procedure is applied to further enhance
the accuracy.

The \textit{Normal Profile Construction Module} constructs the
initial silence profiles based on a short, typically two minutes,
training sample taken when there is no human motion present in the
area of interest. (Section~\ref{sec:nprofile})

The \textit{Basic Detection Module} examines each stream readings in
the monitoring phase and decides whether there is an anomalous
behavior or not. This operation is applied to each stream
independently. It also assigns an anomaly score to each stream to
express the intensity of the anomalous behavior. (Section~\ref{sec:basic})

The \textit{Normal Profile Update Module} updates the normal
profiles constructed in the offline phase in order to adapt to
changes in the environment. (Section~\ref{sec:uprofile})

The \textit{Decision Refinement Module} applies heuristic methods to
refine the decision generated by the basic detection module to
reduce the false alarm rates. (Section~\ref{sec:refine})

The \textit{Region Tracking Interface} provides an interface that
visualizes the output of the above modules. This interface enables
the user to identify the detected events and provides the regions of
the moving entities. (Section~\ref{sec:ui})

We start by giving the mathematical notations followed by the
details of the different modules.

\subsection{Mathematical Notations}

Let $k$ be the number of streams, which is equal to the number of
APs times the number of MPs. Let $s_{j,t}$ denote the received signal
strength (RSS) reading for a stream $j$ that is received at a time instant
$t$. The system studies the behavior of a sliding window $W_{j,t}$
of size $l$ that ends at time $t$, i.e. $W_{j,t} = [s_{j,t-l+1},
s_{j,t-l+2}, ..., s_{j,t}]$.

In order to study the behavior of the sliding windows, each sliding
window $W_{j,t}$ is mapped to a single \emph{feature} or value
$x_{j,t}$ through a function $g$. For example, if the mean is the
selected feature, then $g(W_{j,t}) = \frac{1}{l}\sum_{i=1}^{l}
s_{j,t-l+i}$. Two types of features can be considered: measures of
central tendency, such as the mean, and measures of dispersion or
variation, such as the variance.

\subsection{Normal Profile Construction}
\label{sec:nprofile} 
\begin{figure*}[!t]
    \centering
        \includegraphics[width=0.65\textwidth]{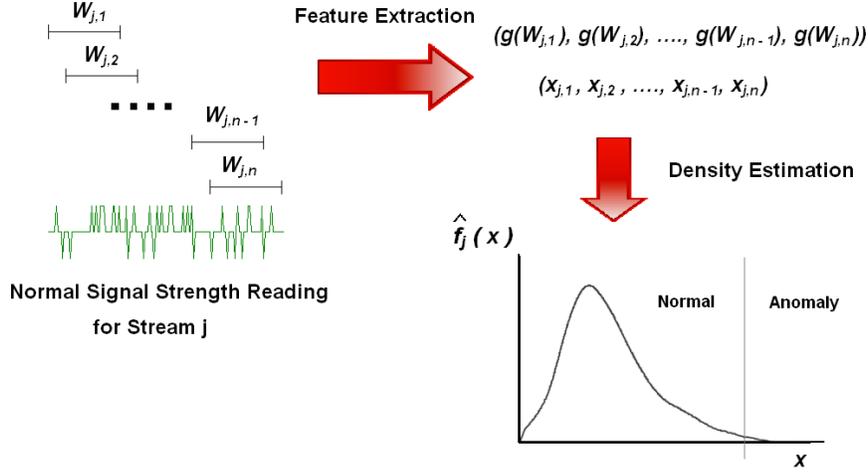}
    \caption{Illustration of the normal profile construction.}
    \label{fig:basic}
\end{figure*}

The purpose of the Normal Profile Construction Module is to construct a
normal profile, capturing the received signal strength characteristics when there is no
human in the area of interest. This is used later by other modules
to detect anomalies.
This module runs in the offline phase. It extracts the feature
values from the sliding windows over the collected data and
estimates its distribution. The density function of the feature
values observed is estimated using non-parametric kernel density estimation\footnote{In Section \ref{sec:analyticalmodel}, we present the motivation for using a non-parametric approach by providing a performance comparison with a parametric modeling of the system operation.}. This
is done for each stream independently. Figure~\ref{fig:basic}
illustrates the operation.

Formally, for a stream $j$, given a set of $n$ sliding windows, each
of length $l$ samples, % (i.e. there are $n+l-1$ readings collected)
each window $W_{j,i}$ is mapped to a value $x_{j,i}$, where $x_{j,i}
= g(W_{j,i})$. Assume $f_{j}$ is the density function representing
the distribution of the observed $x_{j,i}$'s, then given a random
sample $ x_{j,1},x_{j,2},...,x_{j,n}$, the estimated density
function $\hat{f_j}$ is given by
    \cite{Silverman}:
\begin{equation}
\label{eq:kernel_density} \hat{f_j}(x) = \frac{1}{nh_j}
\sum\limits_{i=1}^{n} V \left( \frac{x-x_{j,i}}{h_j} \right)
\end{equation}
where $h_j$ is the bandwidth and $V$ is the
kernel function. The choice of the kernel function is not significant for the results
of the approximation \cite{Scott}. Hence, we choose the Epanechnikov
kernel as it is bounded and efficient to integrate:

\begin{equation}
\label{eq:Epanechnikov} V(q)=
\begin{cases}
 \frac{3}{4} (1-q^2) , & \mbox{if } \left|q\right| \leq 1 \\
 0, & \mbox{otherwise}
\end{cases}
\end{equation}

Also, we used Scott's rule to estimate the optimal
bandwidth \cite{Scott}:
\begin{equation}
\label{eq:Bandwidth} h_j^{\ast}=2.345 \hat{\sigma_j} {n^{-0.2}}
\end{equation}
where $\hat{\sigma_j}$ is an estimate for the standard deviation for
the $x_{j,i}$'s.

After estimating the density function for the feature values
extracted from the sliding windows, critical bounds are selected so
that if the feature values observed in the monitoring state exceed
those bounds, the observed values are considered anomalous.
%The critical values will depend on the type of the feature selected.
Given a significance parameter $\alpha$ and assuming $\hat{F_j}$ is
the CDF of distribution shown in Equation \ref{eq:kernel_density},
if the feature is a measure of central tendency, which can deviate
to the left or the right, then lower and upper bounds will be
calculated such that the lower bound is $\hat{F_j}^{-1}(\alpha/2)$
and the upper bound is $\hat{F_j}^{-1}(1 - \alpha/2)$. However, if
the feature is a measure of dispersion, which can only deviate in
the positive (or right) direction, then an upper bound is only
needed and is equal to $\hat{F_j}^{-1}(1 - \alpha)$.
In the next subsection, we study different features that can be
selected.

\subsection{Feature Selection}
\label{sec:Feature}

As the system requires an offline phase before operation, to learn
the behavior of the signal readings in the normal state, the
selected feature for system operation should be resistant to
possible environmental changes that may affect the stored data, e.g.
temporal variations\footnote{Our experiments show that the changes in the traffic load on the network do not affect the signal strength. Therefore, temporal variations here refer to changes in the physical environment that affect the signal strength.}. In addition, the selected feature should also
be sensitive to the human motion to enhance the detection
accuracy.

In this section, we compare two categories of features: central
tendency measures and dispersion measures. The goal of this study is
to identify the category that will be more promising for the system
operation. For this study, we consider the mean as a central
tendency measure, and the standard deviation as a measure of
dispersion. We use the standard deviation, rather than the variance,
as the variance is a squared measure, while the mean is not.

\subsubsection{Sensitivity to human activity} 
\begin{figure}[!t]
    \centering
        \includegraphics[width=0.5\textwidth]{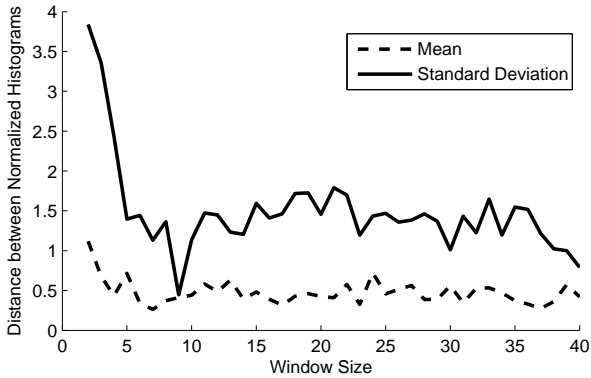}
    \caption{Distance between the features' normalized histograms for silence and motion states.}
    \label{fig:human-effect}
\end{figure}

\begin{figure}[!t]
    \centering
        \includegraphics[width=0.5\textwidth]{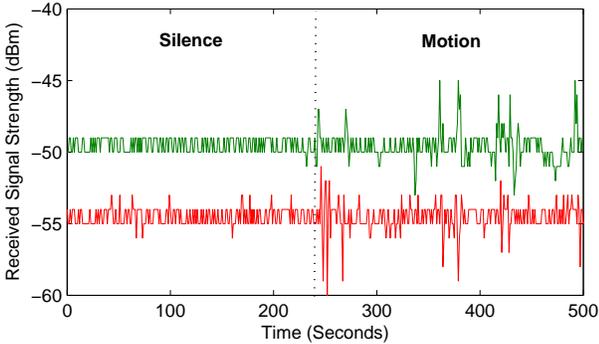}
    \caption{Illustrating the motion effect on wireless signals of two different streams.}
    \label{fig:rss-example}
\end{figure}

The selected
feature should be sensitive to human activity. To compare the two
features, we use the Euclidean distance between the normalized
histograms representing the silence and motion states. The Euclidean
distance is defined as the square root of the sum of the squared
distance between each corresponding histogram bin. The histograms
are constructed over a two-minute period for each state using
Testbed 1, which is discussed later in Section \ref{sec:Evaluation}.
Figure~\ref{fig:human-effect} shows the comparison versus different
window sizes. The figure shows that the distance between the
histograms of the standard deviation is larger than the distance
between the histograms of the mean. This indicates that the standard
deviation feature is more discriminant of the human motion than the
mean feature. This conclusion can be justified by observing the motion effect on
typical wireless signals. Figure \ref{fig:rss-example} provides a visualization of the raw signal strength for two different streams during silence and human motion periods. The figure shows that in the case of human motion, the fluctuations can be up or down around the normal/silence signal level, which leads to a limited effect on the mean as compared to the standard deviation.

 %\comment{comment on shape trend}

\subsubsection{Resistivity to temporal variations} 
\begin{figure}[!t]
        \centering
            \includegraphics[width=0.5\textwidth]{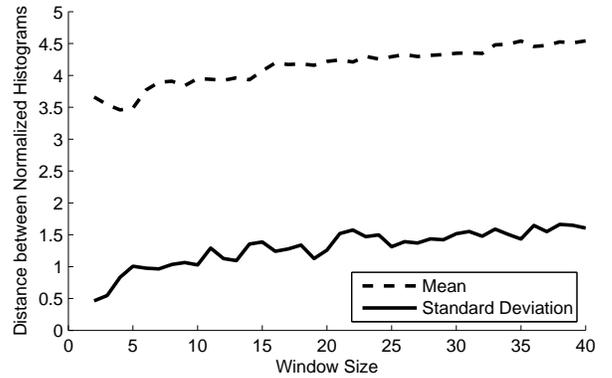}
        \caption{Distance between the features' normalized histograms for the two week-separated silence data.}
        \label{fig:temporal}
    \end{figure}

As the proposed
system requires a learning phase before operation, it is necessary
to reduce the temporal variation effect on the stored profiles. To
compare the two features, we use two different
\emph{\textbf{silence}} data sets collected two weeks apart. Figure~\ref{fig:temporal} shows the results. The more similar the
histograms, the more resistive the feature is to the introduced
variations. The figure shows that the standard deviation feature is
less affected by temporal variations. This is due to the fact that the standard deviation is a relative measure as
it is calculated with respect to the mean, whereas the mean itself provides an absolute value that is more susceptible to be affected by changes in the conditions.

From this study, we conclude that the measures of dispersion, e.g.
the standard deviation or variance, are more suitable for our
proposed system. For the rest of the paper, we use the sample variance as
the selected feature.

\subsection{Basic Detection Procedures}
\label{sec:basic}

The Basic Detection Module runs during the monitoring phase. The purpose of this
module is to detect signal strength anomalies, i.e. human presence,
based on the normal profiles constructed during the offline phase.
In particular, for a window of samples $W_{j,t}$ for stream $j$ at a
given time instant $t$, the module calculates the corresponding
feature value $x_{j,t}$, i.e. the sample variance. A stream $j$ is
considered anomalous if $x_{j,t}$ is above a critical bound $u_j$.
Given a significance parameter $\alpha$ and assuming
$\hat{F_j}$ is the CDF of distribution shown in
Equation~\ref{eq:kernel_density}, the upper bound $u_j$ will be equal to the $100(1-\alpha)^{th}$ percentile of the CDF function, such that $u_{j} = \hat{F_j}^{-1}(1-\alpha)$.

The Basic Detection Module declares a global alarm when any stream is anomalous. This approach can lead to many false positives due to signal strength outliers.  This is enhanced later by the Decision Refinement Module. The Basic Detection Module also calculates an anomaly score $a_{j,t}$ for each stream $j$ to keep track of the significance of any anomalous activity. For a given window, $W_{j,t}$, the anomaly score, $a_{j,t}$, can be calculated as:
$ a_{j,t} = \frac{x_{j,t}}{u_j}$
where $x_{j, t}$ is the sample variance of the window and
$u_j$ is the critical value. This means that a
detected anomaly will have a score greater than one and a silence
window will have a score of less than one. The anomaly score is used by the Normal Profile Update and Decision Refinement modules to further enhance the accuracy.

In summary, the basic detection procedure requires two parameters: the
window size $l$ and the significance $\alpha$. Analysis of both
parameters is presented in Section \ref{sec:Evaluation-performance}.

\subsection{Capturing Changes in the Environment: The Normal Profile Update Module}
\label{sec:uprofile}

Due to the dynamic changes in the environment, the stored profiles
may not capture the real normal state. Therefore, the systems needs
to update the stored profiles during the online phase. The technique
we employ for handling the update process is based on continuously
updating the estimated density in Equation \ref{eq:kernel_density},
by adding $x_{j,t}$'s, that do not have high anomaly scores in
average to it. In particular, during the monitoring phase, the
system groups the consecutive $x_{j,t}$'s in disjoint groups of size
$l_{\textrm{update}}$. The group that has an average anomaly score of less
than one is added to the normal profile. The parameter $l_{\textrm{update}}$
can be tuned to provide the desired performance. We quantify the
effect of the $l_{\textrm{update}}$ parameter in detail in Section
\ref{sec:Evaluation-performance-normal}.

Adding new data to the normal profiles implies the need to give more
weight to the recent data. %Thus, instead of giving equal weights to
%the samples used for the probability calculation in Equation \ref{eq:kernel_density}, more weight will be given to recent data. 
Therefore, Equation \ref{eq:kernel_density} is modified
to:

\begin{equation}
\label{eq:kernel_density2} \hat{f_j}(x) = \frac{1}{h_j}
\sum\limits_{i=1}^{n} w_i V \left( \frac{x-x_{j,i}}{h_j} \right)
\end{equation}
where $\sum\limits_{i=1}^{n} w_i = 1$. We choose linear weights such
that $w_i = \frac{i}{n(n+1)/2}$ ($n$ is constant). We found that exponential weights
do not provide good performance due to the high discrimination
introduced between older and newer data.

\subsection{The Decision Refinement Module}
\label{sec:refine}

Typical wireless environments are noisy. This fact can cause many false alarms if the system generates alarms just based on a single stream. The goal of the Decision Refinement
Module is to reduce the false alarm rate by fusing different streams.

\begin{figure}[!t]
    \centering
        \includegraphics[width=0.5\textwidth]{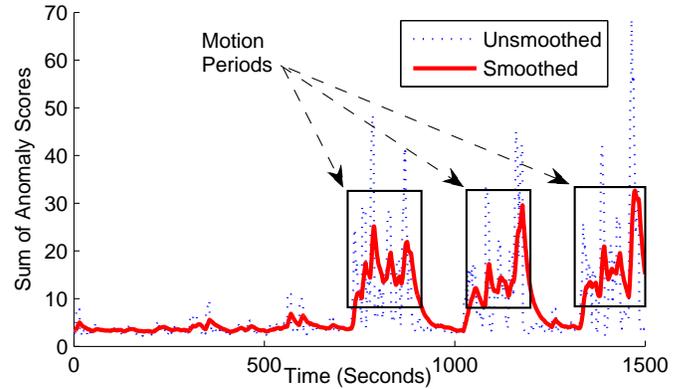}
    \caption{The behavior of the sum of anomaly scores for Testbed 1.}
    \label{fig:sum}
\end{figure}

Since the \emph{Basic Detection Module} assigns an anomaly score to
each detected event that expresses its significance, this can be
leveraged to enhance the detection performance. The Decision Refinement Module studies the behavior of a global anomaly score $a_t$ that is calculated by summing the individual anomaly scores for each stream.
If a noticeable change in $a_t$ occurs, based on a threshold, while at least one stream is
anomalous, this implies the start of an anomalous behavior. The module makes use of the history of the activity state inside the environment through the usage of
\emph{exponential smoothing} to monitor the $a_t$ in order to avoid the
noisy samples, hence reducing the false alarm rate. It also implicitly makes use of the locality of human motion, meaning that the human will continue to affect the same stream and/or other streams near it, causing the sum of anomaly scores smoothed curve to have higher values during the motion period (Figure~\ref{fig:sum}).

\subsection{Region Tracking User Interface Module}
\label{sec:ui} 
\begin{figure}[!t]
  \begin{center}
    \subfigure[Silence State]   {\label{fig:sil-shot}
    \includegraphics[width=0.4\textwidth]{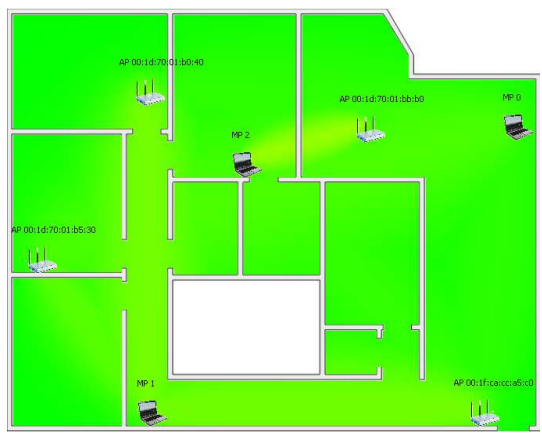}}
    \subfigure[Two Persons Moving] {\label{fig:mov-shot}
    \includegraphics[width=0.4\textwidth]{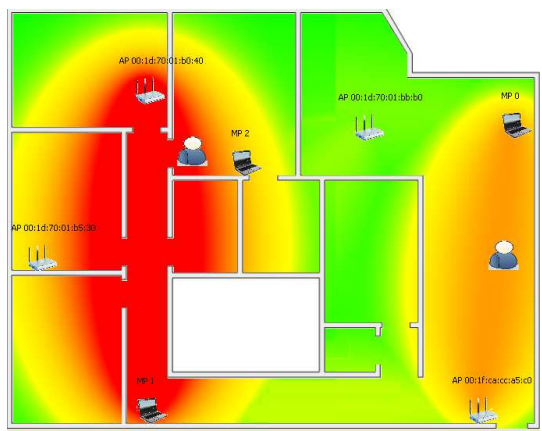}}
  \end{center}
  \caption{The output of the Region Tracking Interface.}
  \label{fig:shot}
\end{figure}

The system provides an interface that provides
information about the probable regions of the detected event. This
is based on visualizing the anomaly degree of each stream enabling
the user to identify the regions that probably have moving entities
inside. This is done by coloring each pixel on the map according to
its distance from each stream endpoints and according to the anomaly
score of each stream. Figure~\ref{fig:shot} displays the output of
this interface when two persons are moving inside a typical site,
showing the true locations of the two persons.

\section{Experimental Evaluation}
\label{sec:Evaluation} In this section, we study the effect of the different parameters on the performance of the \textit{RASID} system and compare it to the previous
WLAN \emph{DfP} detection techniques \cite{Challenges,
Smart}.

\subsection{Experimental Testbeds and Data Collection}
\label{sec:Evaluation-testbeds}

\begin{figure}[!t]
    \centering
        \includegraphics[width=0.5\textwidth]{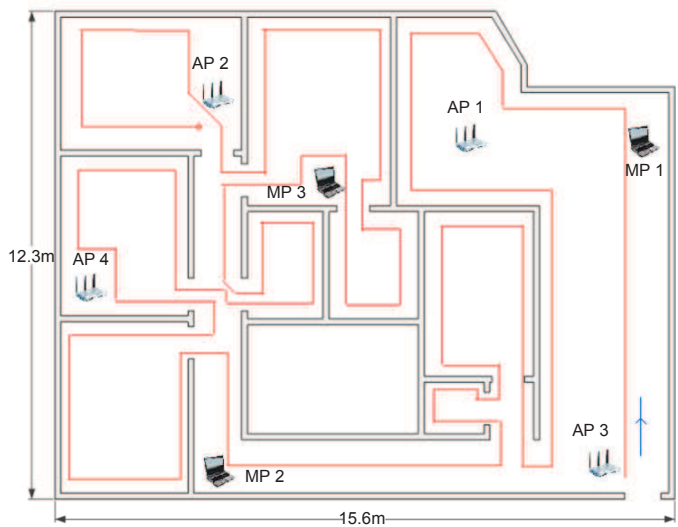}
    \caption{Testbed 1 layout and motion pattern.}
    \label{fig:Exp1}
\end{figure}

\begin{figure}[!t]
  \begin{center}
    \subfigure[Floor 1]   {\label{fig:floor1}
    \includegraphics[width=0.4\textwidth]{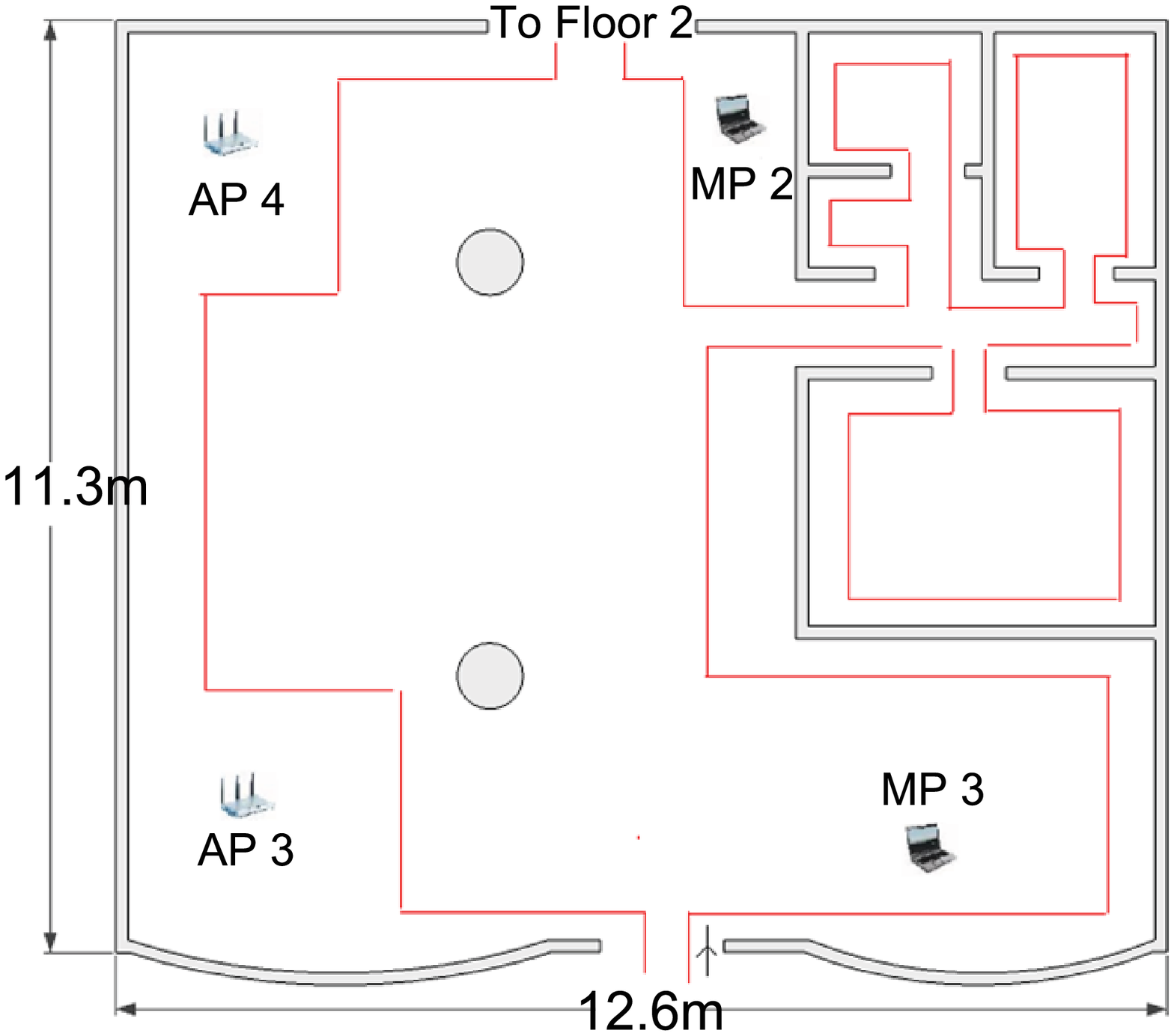}}
    \subfigure[Floor 2] {\label{fig:floor2}
    \includegraphics[width=0.4\textwidth]{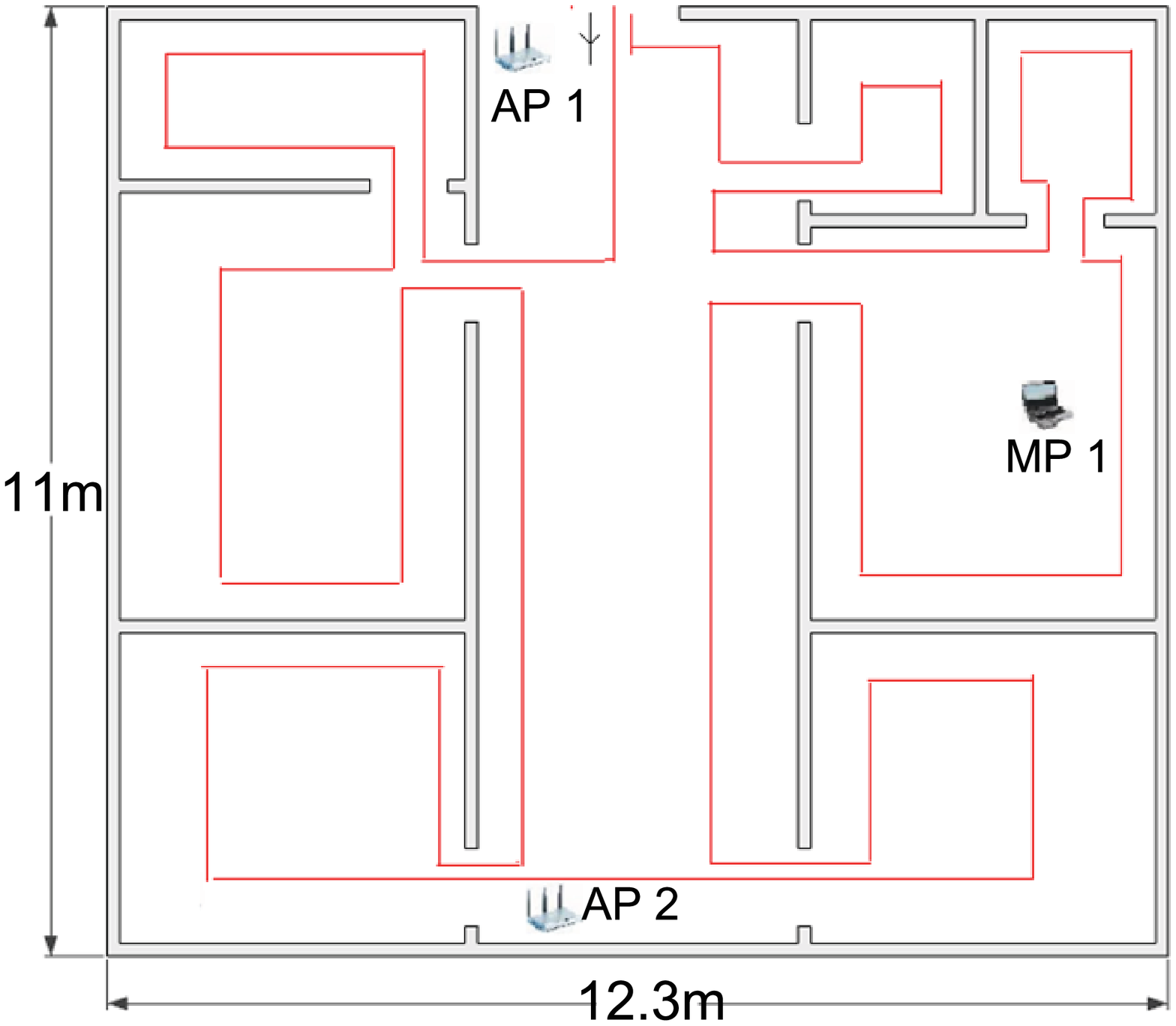}}
  \end{center}
  \caption{Testbed 2 layout and motion patterns.}
  \label{fig:Exp2}
\end{figure}

We collected two sets of data to evaluate the system performance,
each in a different testbed. The first testbed is an
office of approximately 2000 ft$^2$. The second experiment was
conducted in a two-floor home building where each floor was
approximately 1500 ft$^2$. Both tesbeds were covered with typical
furniture. For both testbeds, we used four Cisco Aironet 1130AG
series access points and used three DELL laptops equipped with D-Link
AirPlus G+ DWL-650+ Wireless NICs as MPs. The access points were operating on different channels. The experiments were conducted in
typical IEEE 802.11b environments. Figures \ref{fig:Exp1} and
\ref{fig:Exp2} show the layouts of both experiments.

For the data collection, sets of normal (silence) state readings and
continuous motion readings were collected for each testbed. A total
of about one hour and 15 minutes of data was collected for each
testbed with a sampling rate of one sample per second using the active scanning technique \cite{Horus}. For Testbed 1, this includes three motion sets, while for
Testbed 2, this includes two motion sets. A motion set covers the
entire area of the testbed, as shown in figures
\ref{fig:Exp1} and \ref{fig:Exp2}, and represents the motion of a
single person walking normally around the site without any stops.

For system evaluation, extreme conditions were
employed: The training period is chosen to be only the first
two minutes of the entire data collected with the absence of human
motion. In addition, only one person moved in the area of
interest. More people in the area of interest will lead to higher variance \cite{HumanEffect} and hence better detection. Therefore, the reported results present a lower bound on the performance of the \textit{RASID} system.

\subsection{Evaluation Metrics}
\label{sec:Evaluation-metrics} We used three metrics to analyze the
detection performance: the false positive (FP) rate, the false
negative (FN) rate and the F-measure. The false positive rate refers to the
probability that the system generates an alarm while there is no
human motion in the area of interest. The false negative rate refers
to the probability that the system fails to detect the human motion
in any place in the area. We also use the F-measure, which
provides a single value to measure the effectiveness of the
detection system \cite{IR}.

Since each anomalous sample may not be detected simultaneously, we also studied the detection latency, i.e. how much time the system needs to associate an anomalous sample with a detected event. The overall $90^{th}$ detection latency percentile in both testbeds was found to be less than one second.  

\begin{table*}[t]
    \centering
    		\small
    		
        %\scriptsize
        %\footnotesize
            \begin{tabular}{|r|r|c|c|c|}
            \hline
             &          & Basic Detection & Normal Profile & Decision Refinement\\
             &  				& Module & Update Module  	& Module  (\textbf{RASID Perf.})\\
            \hline
            \multirow{4}{*}{Testbed 1} & FN Rate    & 0.0672 & 0.0876 & \textbf{0.0468}\\
             & FP Rate    & 0.2158 & 0.1176 &\textbf{ 0.0378} \\
             & F-measure   & 0.8683 & 0.8989 & \textbf{0.9574} \\
             & Enhancement & - & 3.52\% & \textbf{10.26}\% \\
            	\hline
            	\hline
            \multirow{4}{*}{Testbed 2} & FN Rate    & 0.2368 & 0.2069 & \textbf{0.0966} \\
             & FP Rate    & 0.1059 & 0.0903 & \textbf{0.0372} \\
             & F-measure & 0.8167 & 0.8422 & \textbf{0.9311} \\
             & Enhancement & - & 3.1\% & \textbf{14}\% \\
             \hline
            \end{tabular}

     \caption{System performance under the same parameters ($l=5, \alpha=0.01, l_{\textrm{update}}=15$) for the two testbeds. Enhancement is with respect to the F-measure of the Basic Detection Module.}
         \label{tab:System Performance}
\end{table*}

\subsection{System Performance}
\label{sec:Evaluation-performance}
Table ~\ref{tab:System Performance} summarizes the system performance for
both testbeds \textbf{\emph{using the same parameters}} for all modules. The table
also shows the enhancement introduced by each module to show the
robustness of the techniques.

\subsubsection{Basic Detection Module}
\label{sec:Evaluation-performance-basic}

\begin{figure}[!t]
%\large
  \begin{center}
    \subfigure[False negative rate]   {\label{fig:FN-3D}
    \includegraphics[width=0.45\textwidth]{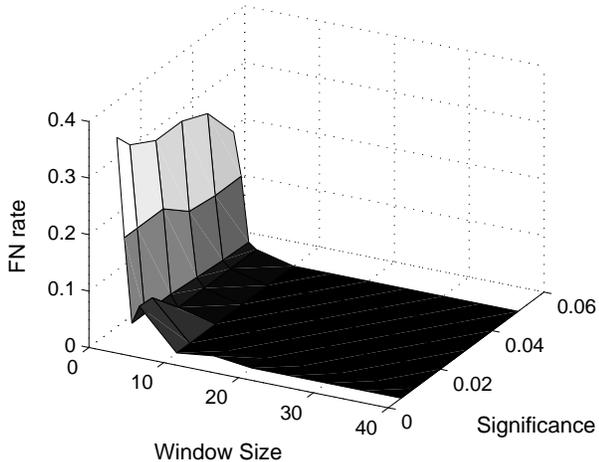}}
    \subfigure[False positive rate] {\label{fig:FP-3D}
    \includegraphics[width=0.45\textwidth]{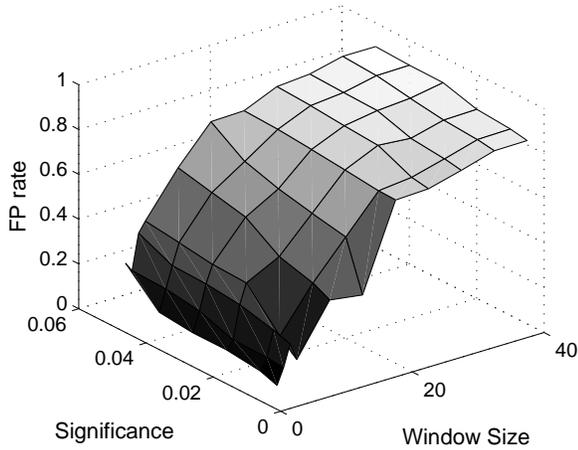}}
%    \subfigure[F-measure] {\label{fig:F-3D}
%    \includegraphics[width=0.3\textwidth]{F-measure.eps}}
    \end{center}
  \caption{Analysis of the Basic Detection Module parameters for Testbed 1.}
  \label{fig:basic-param-analysis}
\end{figure}

As mentioned earlier, this module requires the selection of the
sliding window size $l$ and the significance $\alpha$. Figure~\ref{fig:basic-param-analysis} illustrates the effect of these
parameters applied to Testbed 1. Similar performance has been
observed for Testbed 2.
The figure shows that choosing a too
short window size will make the system less sensitive to human
motion. On the other hand, choosing a very large window size will
introduce a very high FP rate. For the significance parameter, as $\alpha$ decreases, the FP rate
decreases and the FN rate slightly increases. This means that
increasing the significance will result in less system sensitivity.
Therefore, to balance the different performance metrics, we choose
$l = 5$ and $\alpha = 0.01$.

Table~\ref{tab:System Performance} shows that Testbed 2 has a
higher FN rate than Testbed 1 in the Basic Detection Module. This is due to the larger
testbed area (i.e. less coverage) and the time needed to move between the floors in Testbed 2.
This is significantly enhanced by the processing performed by the Normal Profile Update and Decision Refinement modules. It can be noted also that the FP rate in Testbed 1 is relatively high.
This is because the two-minute training period is not enough to sustain accurate
detection for one hour of accurate operation inside the office environment. This highlights the
need for the \emph{Normal Profile Update Module}.

\subsubsection{Normal Profile Update Module}
\label{sec:Evaluation-performance-normal}

\begin{figure}[!t]
    \centering
        \includegraphics[width=0.5\textwidth]{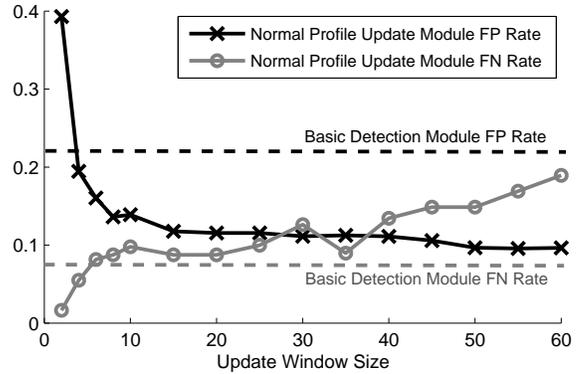}
    \caption{Effect of the update window size parameter ($\l_{update}$).}
    \label{fig:update}
\end{figure}

The Normal Profile Update Module requires the selection of the
update window size $l_{\textrm{update}}$. Choosing a too small $l_{\textrm{update}}$ will
make the system very sensitive to noisy readings causing a high FP
rate. On the other hand, a very large $l_{\textrm{update}}$ will make the system
less sensitive to human motion causing a higher FN rate. Figure~\ref{fig:update} illustrates these effects of the update window size
on the system performance for Testbed 1 when $l = 5$ and $\alpha =
0.01$.
The figure shows that an update window size between 10 and 20 is sufficient to reduce the
high FP rate without causing much increase to the FN Rate.
Thus, we choose $l_{update} = 15$. The
results are shown in Table~\ref{tab:System Performance}. The table
shows that there is about 50\% reduction in the FP rate in the first testbed, but this lead to a slight growth in the FN rate. For Testbed 2, the results of both the FN and the FP rates were enhanced due to adapting to the environment.
Overall, the F-measure was enhanced by 3 to 4\% with respect to the Basic Detection Module performance.

\begin{figure}[!t]
    \centering
        \includegraphics[width=0.5\textwidth]{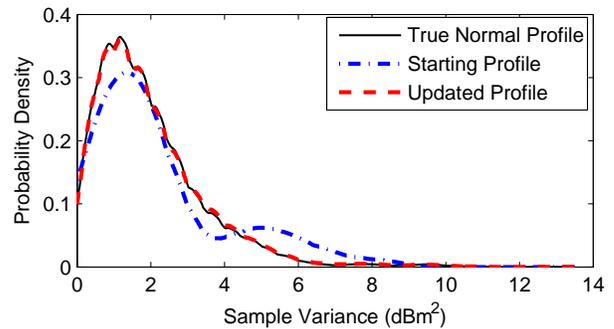}
    \caption{Comparison between the starting profile, updated profile and the true profile for the sample variance of the AP4-MP3 stream at the end of Experiment 1. As shown, the updated and true profiles are almost congruent.}
    \label{fig:update_example}
\end{figure}

This enhancement can be explained by the observation that the Normal Profile Update Module reduces the effect of the temporal variations between the environment true normal profiles and the stored normal profiles by updating them. We verified that by applying the two-sample Kolmogorov-Smirnov test to the distributions of the updated profiles and the distributions of the true normal state. The test accepted the hypotheses that those distributions came from the same underlying distribution at a significance of 0.05. Figure~\ref{fig:update_example} provides an example comparing the starting, updated and true sample variance profiles at the end of Experiment 1.

\subsubsection{Decision Refinement Module}
\label{sec:Evaluation-performance-refinement}

This module fuses the data from all streams. Figure~\ref{fig:sum} displays the sum of anomaly scores curve for the data collected for Testbed 1. To reduce the FP rate, the curve is exponentially smoothed with a smoothing coefficient of 0.04. A large increment in the smoothed curve, by more than 20\% to 25\% from the normal level, implies a period of human motion.  Our experiments show that deviations from these parameters values will not lead to significant degradation in the results.
The figure shows that the motion periods are clearly distinguishable from the silence state. Table~\ref{tab:System Performance} shows that this module can lead to up to 10 to 14\% enhancement in the F-measure for both
testbeds with respect to the Basic Detection Module. It is important to note that this module also reduced the FN rate, as some of the previously undetected events are now detected because this technique makes
use of the history of the state of the activity as described earlier.

\subsection{Comparison with Previous Techniques}
\label{sec:comparison} In this section, we compare the performance
of \textit{RASID} to the previous techniques devised for WLAN
\emph{DfP} detection. We start by a brief description
of the techniques, followed by the different aspects we
evaluate the techniques on. Finally, we present the results
of the comparison.

\subsubsection{Comparison Techniques}
Three techniques are considered for the comparison:
\begin{enumerate}

\item The moving average technique \cite{Challenges}: The moving
average technique uses a\emph{ central tendency feature}, i.e. the
average. It uses two sliding window averages: a short window average
representing the current system condition and a long window average
representing history. The idea is to compare the two averages and if
the difference is above a threshold, a detection is announced. It is
important to note that the moving average technique does not require
a training phase.

\item The moving variance technique \cite{Challenges}: The moving
variance technique uses a \emph{dispersion feature}, i.e. the
variance. Similar to the moving average technique, it compares the
variance of the current system state, based on a sliding window, to
the variance of the silence period, obtained through a training
phase. If the difference is above a threshold, a detection is
announced.

\item The maximum likelihood classification (MLE) technique
\cite{Smart}: This technique constructs profiles for the silence
period as well as for the motions period for different locations in
the area of interest. The profiles represent the signal strength
distribution for each stream at each location. Therefore, it
involves significant training data. During the detection phase, the
system finds the profile that has the maximum likelihood given a
signal strength vector, one entry for each stream. If the estimated
profile corresponds to a motion profile, an alarm is generated.
\end{enumerate}

\subsubsection{Comparison Aspects}

\begin{itemize}
    \item Static accuracy: accuracy when the system is evaluated with the same profiles it was trained on (if any). This is to test the best attainable accuracy.
    \item Profiles' robustness: that is how consistent the performance of the system is when the tested profiles are different from the trained ones, for example due to temporal changes in the environment. For this case, the testing data set is
collected two weeks after the data sets used for training.
    \item Overhead: the effort needed to deploy the system.
\end{itemize}

\begin{table*}%[!t]
    \centering
    \small
            \begin{tabular}{|l|l|l|l|l|l|}
                \hline
                & \multicolumn{5}{|c|}{Results with static profiles}\\
                \cline{2-6}
                 &  & Moving Average & Moving Variance & MLE & \textit{RASID}\\
                \hline
                \multirow{3}{*}{Testbed 1} & FN Rate & 0.1446 & 0.1426 & 0.0363 & 0.0468\\
                 & FP Rate & 0.1385 & 0.104 & 0.1547 & 0.0378\\
                 & F-measure & 0.858 & 0.8743 & 0.9099 & \textbf{0.9574}\\
                \hline
                \multirow{3}{*}{Testbed 2} & FN Rate & 0.0759 & 0.308 & 0.0372 & 0.0966\\
                 & FP Rate & 0.7412 & 0.1478 & 0.0774 & 0.0372\\
                 & F-measure & 0.6935 & 0.7522 & \textbf{0.9438} & 0.9311\\
                \hline
                \multirow{3}{*}{} & Overhead & No overhead & Minimal & Worst & Minimal \\
                \hline
                \hline
                & \multicolumn{5}{|c|}{Results with testing profiles separated two weeks}\\
                & \multicolumn{5}{|c|}{from the training profiles.}\\
                \cline{2-6}
                 &  & Moving Average & Moving Variance & MLE & \textit{RASID}\\
                \hline
                \multirow{3}{*}{Testbed 1} & FN Rate & 0.2165 &0.319 &  0.1653 & 0.0472\\
                 & FP Rate & 0.0711  &0.1561  & 0.952 &  0.0782\\
                 & F-measure & 0.8449& 0.7414 & 0.5991 & \textbf{0.9383}\\
                \hline
                \multirow{3}{*}{Testbed 2} & FN Rate & 0.2641 & 0.4152 & 0.1203 &  0.0931\\
                 & FP Rate & 0.3602 &0.0513 &  0.831  & 0.0722\\
                 & F-measure & 0.7022 & 0.7149& 0.6491& \textbf{0.9165}\\
                \hline
                \multirow{3}{*}{} & Overhead & No overhead & Minimal & Worst & Minimal \\
                \hline
        \end{tabular}

            \caption{Performance comparison with previous \textit{DfP} detection techniques.}
                    \label{tab:comparison}
\end{table*}

\subsubsection{Comparison Results} Table~\ref{tab:comparison} shows the comparison results in two cases.

In terms of the static accuracy, the results show that the F-measure of the \textit{RASID} system is better than
other systems in Testbed~1 and is slightly lower than the
MLE technique in Testbed~2. Compared to the Moving Average and Moving Variance techniques, the \textit{RASID}
system provides high accuracy due to the techniques it uses to enhance the performance. On
the other hand, the MLE technique achieves slightly
higher accuracy in Testbed~2 as it stores a motion profile, which
requires much higher overhead than the \textit{RASID}
system.

In terms of profiles' robustness,  the Moving Average technique does not store any profiles. Therefore, its
overall performance is low but almost the same as the profiles
change. On the other hand, the robustness of the MLE technique is
the least as it uses the mean signal strength values
as the features used for classification. Therefore, after two weeks,
the distribution of the signal strength does not follow the learned one. This is why the FP rate for
the MLE technique is too high due to the shift that occurred in the
signal distributions. It can also be noted that
\textit{RASID} performance in the two cases was the best
because \textit{RASID} uses the variance for its operation (dispersion feature) and employs techniques for adapting to changes in the environment and for enhancing the performance. This is why \textit{RASID}
performance is better than the Moving Variance in general, although
the Moving Variance uses the same feature as \textit{RASID}.
%

%\item
In terms of overhead, the Moving Average
technique has the minimum overhead as it does not need any learning
phase. The Moving Variance and \textit{RASID} deployment need to
construct normal profiles by collecting samples for two
minutes when the human is not present. On the other hand, the
MLE technique has the worst overhead
as it constructs motion profile at each location in the
area of interest in addition to the normal profile.
%\end{enumerate}

In summary, although the static detection accuracy of \textit{RASID}
is as accurate as the MLE technique, the MLE technique has significantly higher
overhead than \textit{RASID} because of its motion profile
requirements. In addition, \textit{RASID} is the most robust technique to temporal changes in
the training profiles and significantly outperforms the remaining techniques.

\section{Comparison with a Parametric Approach}
\label{sec:analyticalmodel}
In this section, we compare the performance of the system's non-parametric approach to an analytical model that models the sample variance parametrically. The results of this model can help validate the results of our parameter analysis in the previous section and can also motivate the usage of the non-parametric density estimation. The next evaluation will be based on the results of the Basic Detection Module only, so as to evaluate the two approaches without the enhancements. First, we describe the analytical model, then we present the results of the comparison.

\subsection{The parametric model}
The sample variance can be modeled parametrically given some conditions. According to Cochran's Theorem \cite{Cochran}, the sample variance of $l$ independent normally distributed random samples follows a chi-square distribution with $l - 1$ degrees of freedom such that $\frac{(l-1)s^2}{\sigma^2} \sim \chi^2_{l - 1}$, where $\sigma^2$ is the population variance.
According to \cite{model}, the signal strength readings (in dBm) distributions for a stream $j$ can be assumed to follow a normal distribution.
Given that assumption, a parametric model can be devised for the system when the sample variance is used. Given a significance $\alpha$ and a window size $l$, the upper bound for the sample variance observed during the monitoring phase is $ \chi^2_{l - 1, 1 - \alpha}$. However, \cite{model} also stated that the normality assumption may not hold in some cases. In addition, the signal strength readings may not be independent \cite{correlation}. Thus, we believe that the non-parametric model described before will provide better performance than the parametric model. This will be verified in the following subsection.

\subsection{Analysis Results}

\begin{figure}[!t]
    \centering
        \includegraphics[width=0.5\textwidth]{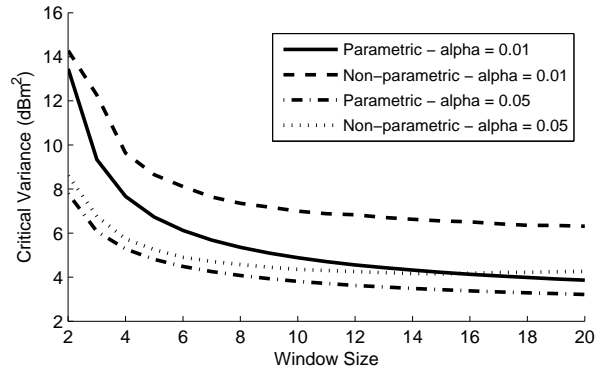}
    \caption{Comparison of the critical variance values of the parametric model and the \emph{RASID} system model (non-parametric approach).}
    \label{fig:critical-variance-comparison}
\end{figure}

First, to check how close the parametric model is to the actual system, we compare the critical upper bounds obtained by both methods. For example, in Figure~\ref{fig:critical-variance-comparison} we compare the critical sample variance values in both cases for the stream AP4-MP3 from Experiment 1, when the population variance is assumed to be 2.02 dBm$^2$ which is an experimental estimate for the population variance of that stream. The figure shows that the parametric model and the actual system critical values follow the same trends. However the difference between the curves suggests that the real case does not exactly follow the assumed parametric model. In addition, the effects of Basic Detection Module parameters can be inferred from the parametric model curves. As the window size parameter $l$ increases, the critical variance value decreases which results in increased system sensitivity (i.e. higher FP rate and lower FN rate). Also, as the significance parameter $\alpha$ increases, the critical variance value decreases which also results in increased sensitivity. This is consistent with the analysis presented in Figure~ \ref{fig:basic-param-analysis}.

\begin{figure}[!t]
    \centering
        \includegraphics[width=0.5\textwidth]{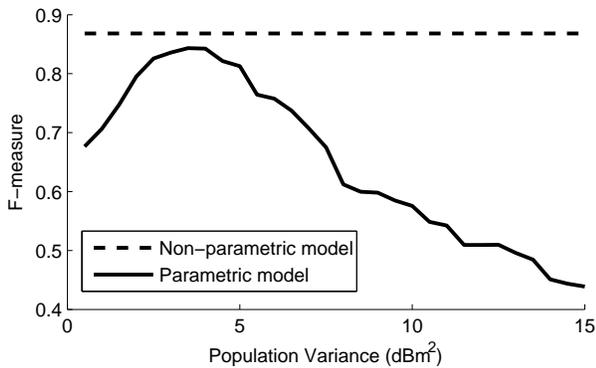}
    \caption{The performance of the Basic Detection Module when the parametric model is used versus the population variance, given $l = 5$ and $\alpha = 0.01$. The best population variance configuration provides an F-measure of 0.843 compared to 0.8683 that was obtained using the non-parametric estimation.}
    \label{fig:model-performance}
\end{figure}

The next point is to study how the usage of the parametric model instead of non-parametric estimation can affect the system performance. As the distribution of the sample variance depends on the population variance, we analyze its effect. Figure~\ref{fig:model-performance} shows the effect of the population variance on the performance of the Basic Detection Module when the parametric model is used for Experiment 1. From the figure, we can conclude that the best performance achieved in terms of the F-measure (0.843) is less than the F-measure obtained using non-parametric estimation (0.8683).

To conclude, the parametric model leads to lower performance compared to the non-parametric estimation because the assumptions that the signal strength values are independent and follow a normal distribution may not hold. Also, the parametric model requires the selection of an accurate population variance. This cannot be done accurately without training for long time periods. Therefore, we conclude that \emph{RASID} approach of constructing non-parametric profiles in a short offline phase and updating them in the online phase does provide a better option.

\section{Discussion}
\label{sec:Discussion}

In this section, we discuss some points related to the configuration
and the performance of the \textit{RASID} system. We also highlight
some research issues and some challenges that can be addressed in
future work.

\subsection{Univariate VS Multivariate Density Estimation}
As mentioned before, the basic detection module studies each stream
independently by estimating the univariate density for the selected
feature of the sliding windows extracted from the training data.
Another possibility was to construct a multivariate density estimate
for the data of all streams. This implies a modification to the
anomaly detection criteria. Different algorithms can be applied in
this case, e.g. \cite{ParzenDetectors}. Our experience with this
algorithm shows that this leads to a degradation of the system
accuracy. The main reason for this degradation is that the system
sensitivity is significantly reduced, especially when the number of
streams is large. In that case, the system may not be able to detect
an anomaly in one stream only, as its effect may not be much sensed.

\subsection{Effect of Network Activity on System Profiles}
\label{sec:nw_activity_effect}
Typically in real wireless environments, it is expected that many monitoring points may be using the wireless network for handling typical tasks (e.g. downloading updates or patches). The question is whether such network activities will require any change in the system normal profiles if they were originally collected while there is no network activity. In this subsection, we present an experimental study to investigate that effect.

\begin{figure}[!t]
    \centering
        \includegraphics[width=0.5\textwidth]{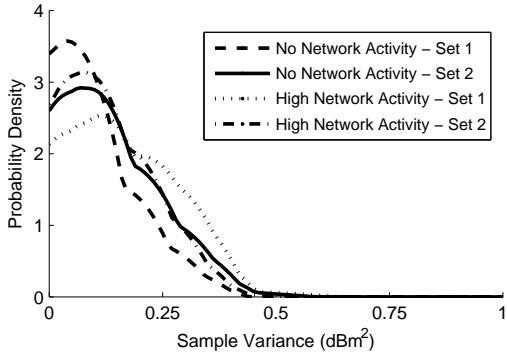}
    \caption{The effect of network activity on the constructed sample variance normal profiles. As shown, the difference between the profiles is not significant.}
    \label{fig:network-activity}
\end{figure}

In order to examine that effect, a simple experiment was conducted on a single stream between an access point and a laptop acting as a monitoring point in silence state. Two signal strength data sets were collected while there was no network activity at the monitoring point, while another two sets were collected while the monitoring point were downloading data through the wireless stream with the maximum download speed allowed (50 KBytes per second). The collected data are used to construct normal profiles in the same way presented earlier in Section \ref{sec:nprofile}. Figure \ref{fig:network-activity} compares the constructed profiles for the four sets. From the figure, it is clear that the difference between the distributions in both cases is negligible. Furthermore, we apply the two-sample Kolmogorov-Smirnov test to each of the four different pairs of those constructed profiles. The test accepted the hypotheses that those estimated distributions came from the same underlying distribution with a significance of 0.05. Therefore, we can conclude that the constructed sample variance profiles are invariant with respect to the state of network activity. 

\subsection{Detection and Identification of Independent Events}

The above experiments showed that the system is capable of detecting
a single person moving inside the area of interest. Obviously, the
detection performance will be enhanced if there were more than one
entity in the area of interest. We verified that the system will be
able to declare that there is anomalous behavior inside the area
more clearly in this case.

It would be useful to \emph{identify} the number of moving entities
in some applications. Figure \ref{fig:shot} shows that we can detect
that there are multiple entities in the area of interest. However,
as our system uses limited data to satisfy the feasibility design
goal (normal profiles only), the system cannot provide full
information about the number of entities in all cases. For example,
if two entities are affecting a single stream only, the system will
detect them as one entity. This is because there is no enough
information that enables the system to differentiate between the two
cases. On the other hand, in some cases, the system can tell with
high probability that some events are due to independent entities.
Here, we briefly describe the constraints through which the system
can provide information about the number of independent entities.

First, let $T_{min}$, a $k \times k$ square matrix denote what we
call a minimum time reachability matrix. Each entry in this matrix
stores the minimum time needed for an entity to affect two streams
$i$ and $j$, such that

\begin{equation}
\label{min_time} T_{min_{ij}} = \frac{D_{min_{ij}}}{v_{max}}
\end{equation}
, where $D_{min_{ij}}$ represents the minimum distance between the
nearest two points on the $i$ and $j$ streams lines of sight and
$v_{max}$ represents the maximum movement velocity inside the area.
The distance $D_{min_{ij}}$ can be calculated from the site map, and
$v_{max}$ can be estimated based on empirical observations.

Two events $E_1$ and $E_2$ are considered independent (i.e. not
generated by the same entity), if they satisfy the following
conditions. First, they should be affecting two different streams
$i$ and $j$ and second, the time difference between $E_1$ and $E_2$
is less than the value $T_{min_{ij}}$. The time difference between
the two events are calculated based on the time difference between
the times when the anomaly scores for the two events reach the peaks
as they express the moments when the entities are affecting the
streams the most. To tell that $n$ events are independent, each pair
of those events should satisfy the conditions described above. The
above conditions imply that the system cannot detect more than $k$
moving entities, where $k$ is the number of streams as stated
earlier.

To conclude, despite the limited information the system uses, the
system can provide information about the number of independent
events inside the monitored area given some conditions. The
significance of this point can be clear when applied inside large
scale environments.

Another possibility is to use the level of the change in variance as
an indication of the number of entities. The hypothesis is that the
more human affecting a single stream, the higher the variance should
be. This hypothesis still needs to be verified though.

\subsection{Integration with DfP Tracking Systems}
Our system can provide useful information to \textit{DfP} tracking
systems like the ones proposed in \cite{Analysis, Nuzzer}. First, a
\textit{DfP} tracking system can use our system to decide whether to
start the tracking process or not. Also, the system can enhance the
tracking accuracy by limiting the probable locations to a certain
area (e.g. as in Figure \ref{fig:shot}). In addition, given the
conditions described earlier, our system can help the tracking
system identify the number of intruders and the area of each one, so
that it can apply the tracking algorithms to each area
independently. This will need further investigation and
experimentation.

\subsection{Combining Features}
Although we showed in this paper that using the variance as a
feature is better than using the mean, both features can be used
concurrently to achieve better performance. Our initial results show
that combining both features and using a simple voting scheme can
enhance the results in some cases. This is a subject for future
investigations.
\subsection{Signal Strength Readings Synchronization}
The synchronization of the signal strength readings received at the
monitoring point can be necessary in some cases. For example, the
technique described before for checking the independence of the
detected events requires synchronization of the readings across the
streams. In addition, the decision refinement module requires the
different streams to be synchronized. In this paper, we took a
centralized approach for synchronization, where the application
server requests the MPs to initiate a reading. Other approaches,
such as time synchronization of the MPs can be employed. The
advantages and disadvantages of each technique in terms of accuracy
and overhead can also be investigated.
\subsection{Effect of Different Hardware}

The hardware used to capture the signal strength values can affect
system performance. Through our experiments, we studied how the WLAN
NIC type affects the quality of the collected readings. We found
that NICs differ in two main aspects: sensitivity to human activity
and noise readings. For example, some cards cannot sense the human
shadowing effect unless it is sustained for a sufficient period of
time. The readings of some other cards are noisy and requires
extensive filtering. These experiments considered the NICs only.
However this can hold from the APs perspective too. Therefore, we
believe a study is needed to identify which hardware will be more
suitable for the system operation and how to account for these
variation between cards and allow the system to operate with
different cards.

\section{Conclusions and Future Work}
\label{sec:Conclusion} In this paper, we presented the
\textit{RASID} system, a system that enables device-free passive
motion detection using the already installed wireless networks.
\textit{RASID} uses non-parametric statistical anomaly detection techniques to
provide the detection capability. The \textit{RASID} system
also employs profile update techniques to capture changes in the
environment and to enhance the detection accuracy. The system was evaluated in two different real environments.
Using the same parameters for the two testbeds, the system provided an
accurate detection capability reaching an F-measure of at least 0.93 in both testbeds.
The performance of the \textit{RASID} system was compared to the previously introduced techniques for WLAN
\textit{DfP} detection. The results showed that the \textit{RASID} system outperformed the state-of-the-art techniques in terms of robustness and accuracy. In addition, we showed that the non-parametric approach employed by \emph{RASID} has significant advantages over a parametric approach for the system operation. 

Currently, we are expanding \emph{RASID} in several directions: One direction is to integrate \textit{RASID}'s detection capability with \textit{DfP} tracking systems while considering larger testbeds. Another direction is to study possible sources of noise in typical wireless environments, e.g. other devices inside or outside the area of interest, and how to reduce their effect. We are also studying how the detected entity's characteristics, e.g. size, shape and motion pattern, can affect the system performance. Moreover, the site configuration, i.e. the positions of the APs and MPs, can also be studied in order to optimize the system performance.

%\bibliographystyle{IEEEtran}
%\bibliography{IEEEabrv,ref}


\end{document}